\documentclass[german,11pt]{article}
\usepackage{graphics}
\begin{document}
\pagenumbering{arabic}

\noindent

%*****
\renewcommand{\baselinestretch}{2}

\noindent
%Letter to Nature\\

\medskip

\noindent
{\Huge
Discovery of molecular hydrogen in a \\
\\
\\
high-velocity cloud
of the Galactic halo}

\bigskip

\noindent
\\
{\small P. Richter $\ast$ , K.S. de Boer $\ast$,
H. Widmann $\dagger$, N. Kappelmann $\dagger$,
W. Gringel $\dagger$,
M. Grewing $\dagger$\\
\\
\noindent
\& J. Barnstedt $\dagger$}
\\
\noindent
\vspace{0.2cm}
\\
{\small $\ast$ Sternwarte der Universit\"at Bonn, Auf dem H\"ugel 71,
D-53121 Bonn, Germany}\\
\\
\noindent
{\small $\dagger$ Institut f\"ur Astronomie und Astrophysik,
Universit\"at T\"ubingen, Waldh\"auserstr. 64, 
\\
\\
\noindent
D-72076 T\"ubingen, Germany} \\
\noindent

\small\normalsize
\vspace{0.5cm}

The Galactic halo contains in many directions
clouds of neutral hydrogen with high radial velocities$^{1}$.
These high-velocity clouds mostly have motions
towards the Galactic plane but do not follow the
rotational movement of the disc.
Since distance information for high-velocity clouds
is sparse$^{2,3}$ and the kinematics cannot be fixed from just
the radial velocites, a closed theory about the
origin of high-velocity clouds is not available today. 
One idea is that
these clouds are the cooling part of a Galactic fountain$^{4,5}$.
Here the gas would predominantly originate in
the metal-rich Galactic disc. The other suggests that they
represent metal-poor gas falling onto the Milky Way from 
intergalactic space$^{6,7}$. The presence of molecular hydrogen, 
whose formation requires that
dust and thus metals are present,
might discriminate between the two theories. 
Here we report the discovery of
molecular hydrogen in a Galactic
high-velocity cloud in the southern sky.
For the same cloud we derive an iron abundance
which is half of the solar value. 
Thus, all evidence points to a Galactic origin for
this high-velocity cloud.

\bigskip
\bigskip

An earlier search for molecular
gas in the Galactic halo 
was limited to radio observations of the
carbon-monoxide molecule (CO)$^{8,22}$.
The by far most abundant molecule in the interstellar 
medium, molecular hydrogen (H$_2$), 
can in its cold form only be
observed in the far ultraviolet (FUV)
domain in absorption against bright 
background sources.
This method requires a telescope working
outside of the earth's atmosphere. 
The echelle spectrograph of the Orbiting and Retrievable Far and 
Extreme Ultraviolet Spectrometer (ORFEUS)$^{9}$
is the first instrument capable to 
investigate the H$_2$ molecule in the FUV outside
the Milky Way disc. Previous instruments
could not match ORFEUS because of
their wavelength coverage or the
lack of sensitivity.

For the investigation of the H$_2$ absorption
in Galactic high-velocity clouds (HVCs)
several lines of sight towards bright stars
in the Magellanic Clouds are available from
ORFEUS.
The two Magellanic Clouds are the most nearby
satellite galaxies of the Milky Way.
The HVC in
front of the 50\,kpc distant Large Magellanic Cloud (LMC) at
$v_{\rm rad} = +120$\,km\,s$^{-1}$ is seen in
almost all absorption spectra of LMC stars$^{10,11}$.
That HVC thus was known to have a substantial extent in the
sky$^{12}$ and therefore most likely is a foreground
cloud of the Milky Way halo.

From the four ORFEUS spectra of LMC stars
the one to the B type star
HD\,269546 ($l=279.32$, $b=-32.77$)
shows the presence of H$_2$
at $+120$\,km\,s$^{-1}$.
This line of sight is the one in our sample with the
highest column density of atomic hydrogen$^{13}$.
For all other LMC spectra no clear absorption is seen
at HVC velocites, probably a result from a combination of
lower H\,{\sc i} column densities, lower signal-to-noise ratios
and line blending.
Nine H$_2$ absorption lines
from the lowest 7 rotational states
and free from line blends were found in 
the spectrum of HD\,269546 and
for 5 additional transitions upper limits
for the equivalent widths were determined
(Table 1).
Fig.\,1 shows part of the spectrum
where some of the detected
H$_2$ absorption features are found.
Fig.\,2 shows the Werner Q(2),0-0 
line at 1010.941 \AA, plotted in velocity
scale as an example for the H$_2$ absorption 
pattern. The H$_2$ absorption at $v_{\rm rad} = +120$\,km\,s$^{-1}$
matches perfectly the neutral hydrogen emission
profile of this sight line$^{13}$ and there is no doubt
that both neutral and molecular hydrogen belong to
this HVC.

For the quantitative analysis we made use of a
standard curve-of-growth technique.
We measured equivalent widths for the H$_2$ absorption
at +120 km\,s$^{-1}$ and constructed
curves of growth for each roational state
$J$ individually which collectively fit the
velocity-dispersion parameter $b = 9$ km\,s$^{-1}$.
The high $b$-value
indicates that the gas is turbulent.
We find the H$_2$ column densities $N(J)$ to be as:
log\,$N(0) \le 14.2$, log\,$N(1) = 14.6 \pm 0.2$,
log\,$N(2) = 14.8 \pm 0.2$, log\,$N(3) = 14.8 \pm 0.2$, 
log\,$N(4) = 14.7 \pm 0.4$,
log\,$N(5) \le 14.7$, log\,$N(6) \le 14.9$.

The total H$_2$ column density in this HVC is small, 
$N($H$_2) = 2.2-3.6 \times 10^{15}$ cm$^{-2}$. 
Comparison with the column density of neutral hydrogen
$N($H{\sc i}$) = 1.2 \times 10^{19}$ cm$^{-2}$, 
derived from the older Parkes 21-cm emission profile$^{13}$,
shows that the proportion of H$_2$ to H\,{\sc i} is similar to that found 
in diffuse gas in the Galactic disc$^{14}$. 
It is indicative of well illuminated gas with low dust content. 

The distribution of column densities over the rotational states 
can be used to roughly estimate the excitation
conditions in the H$_2$ gas$^{15}$. 
The temperature of the H$_2$ gas is $\ge 1000$\,K,
derived by a fit of the population of the rotational levels
to a theoretical Boltzmann distribution. 
This value most likely is not the actual kinetic
temperature, but represents the 
excitation through UV photons from the Galactic disc.
In sharp contrast, the 
relative abundance of para-H$_2$ (even $J$ numbers)
and ortho-H$_2$ (odd $J$ numbers) for $J \le 3$ is 
$\approx 1$. Thus the ortho-to-para ratio must
have been thermalized by proton exchange$^{16}$ with H$^+$ 
to a gas temperature below $100$\,K.
The discrepancy between these two temperature ranges is a 
sign that the gas is {\it not} in thermodynamical equilibrium.
Further support for this is given by the fact that the
observed low H$_2$ column density can hardly be described with approved
theoretical models of H$_2$ formation- and destruction processes$^{17}$ in regard to
the strength of the radiation field from the Milky Way disc.

Recent H\,{\sc i} data from Parkes$^{18}$ 
(Fig.\,3) show that our sight-line crosses the HVC at $+120$\,km\,s$^{-1}$
in a region between two denser cores in a much larger complex.
It shows spatial similarities to Chain\,A,
recently found to be at a distance between $2.5$ and $7$ kpc$^{3}$.
In view of the structure
of the HVC it seems plausible that the cloud core 
recently has fallen apart.
In that case, cold H$_2$ from the cloud core 
was 
exposed
to the unshielded UV flux from the Galactic disc that
currently photodissociates the H$_2$ gas.

In order to further explore the origin of the HVC in front of the
Magellanic Clouds we have combined ORFEUS data with 
public data from the International Ultraviolet
Explorer (IUE) and have determined the iron abundance
in the HVC component at $+120$ km\,s$^{-1}$.
From an analysis of five Fe\,{\sc ii} absorption
lines we derive $b=9$ km\,s$^{-1}$ and 
a column density of log $N($Fe\,{\sc ii}$)
= 14.3 \pm 0.2$, in good agreement
with an earlier investigation$^{19}$.
The $b$-value is the same as for H$_2$, supporting
it is the same gas. The iron abundance is 
log\,$[$Fe\,{\sc ii}/H$] = -4.8$,
which is half of the solar value$^{20}$.
This ratio is based on $N($H\,{\sc i}$)$ from the older
15$^{\prime}$  beam Parkes measurement.
In view of this high metallicity we can rule out that
the gas, and thus the H$_2$, 
is of primordial origin or from intergalactic
space$^{23}$. 

Our results favour the Galactic fountain model$^{4,5}$ 
as explanation for the origin of the HVC gas in front 
of the LMC. In this model hot
gas is ejected out of the Galactic
disc, cools down and 
falls back onto the Galactic plane. 
The molecular hydrogen then must have been formed {\it within} the
cooled halo cloud and at least a small amount of dust must
be present in that HVC.

Theoretical considerations$^{8}$ show that molecular 
hydrogen can form in HVCs once the hydrogen volume density $n$(H\,{\sc i})
exceeds a critival value. This critical volume density 
depends on the H\,{\sc i} column density, the amount of dust,
the radiation field, the gas temperature and
the velocity dispersion parameter $b$. We have
found 
$b=9$\,km\,s$^{-1}$ for our HVC gas, indicating that
H$_2$ might form at volume densities as low as $n$(H\,{\sc i})$
\approx 10$\,cm$^{-3}$. The {\it actual} volume
density can only be determined assuming a certain distance
to the HVC in combination with assumptions for the
depth-structure. However, the uncertainties for
these parameters are too high to judge whether 
the theoretical models can derscribe the observations
we have presented here.

\newpage

\noindent
{\bf References}
\\
\noindent
$^{1}$ Wakker B. Distribution and origin of high-velocity clouds.
{\it Astron.Astrophys.}\,{\bf 250},499-508\,(1991).\\
$^{2}$ Wakker B.P., van\,Woerden H. High-velocity clouds.
{\it Ann.Rev.Astron.Astrophys.}\,{\bf 35}, 217-266\,(1997).\\
$^{3}$ van\,Woerden H. {\it et al.} A confirmed location in the 
Galactic halo for the high-velocity cloud 'chain A'.
{\it Nature}\,{\bf 400}, 138-141\,(1999).\\
$^{4}$ Shapiro P.R., Field G.B. Consequences of a new hot component of the
interstellar medium.\\
{\it Astrophys.J.}\,{\bf 205}, 762-765 \,(1976).\\
$^{5}$ Bregman J.N. The galactic fountain of high-velocity clouds.
{\it Astrophys.J.}\,{\bf 236}, 577-591\,(1980).\\
$^{6}$ Oort J.H. The formation of galaxies and the origin of the
high-velocity hydrogen.\\
{\it Astron.Astrophys.}\,{\bf 7}, 381-404 \,(1970).\\
$^{7}$ Blitz L., {\it et al.} High-Velocity Clouds: Building Blocks of the Local
Group.
{\it Astrophys.J.}\,{\bf 514}, 818-843 \,(1999).\\
$^{8}$ Wakker B.P., Murphy E.M., van Woerden H., Dame T.M.
A sensitive search for molecular gas in high-velocity clouds. {\it Astrophys.J.}
\,{\bf 488}, 216-223\,(1997).\\
$^{9}$ Barnstedt J., {\it et al.} 
The ORFEUS II Echelle Spectrometer: Instrument description, performance and data
 reduction.
{\it Astron.Astrophys.Suppl.}\,{\bf 134}, 561-567\,(1999).\\
$^{10}$ Savage B.D., de Boer K.S. Ultraviolet absorption by interstellar gas at
 large distances from the
Galactic plane. {\it Astrophys.J.}\,{\bf 243}, 460-484\,(1981).\\
$^{11}$ Bomans D.J., de Boer K.S., Koornneef J., Grebel E.K. C\,IV absorption from
 hot gas inside the supergiant
shell LMC4 observed with HST and IUE.
{\it Astron.Astrophys.}\,{\bf 313}, 101-112\,(1996).\\
$^{12}$ de\,Boer K.S., Morras R., Bajaja E. The location of intermediate- and 
high-velocity gas in the general direction of the Large Magellanic Cloud.
{\it Astron.Astrophys.}\,{\bf 233},523-526\,(1990).\\
$^{13}$ McGee R.X., Newton L.M., Neutral hydrogen in the Galactic halo.\\
{\it Proc.Astron.Soc.Australia}\,{\bf Vol.6, 3}, 358-385\,(1986).\\
$^{14}$ Savage B.D., Bohlin R.C., Drake J.F., Budich W. A survey of interstellar
molecular hydrogen.
{\it Astrophys.J.}\,{\bf 216}, 291-307\,(1977).\\
$^{15}$ Spitzer L., Zweibel E.G. On the theory of H$_2$ rotational excitation.
{\it Astrophys.J.}\,{\bf 191}, L127-L130\,(1974).\\
$^{16}$ Burton M.G., Hollenbach D.J., Thielens A.G.G.M. Mid-infrared rotational
 line emission from
interstellar molecular hydrogen.
{\it Astrophys.J.}\,{\bf 399}, 563-572 \,(1992).\\
$^{17}$ Jura M.A. Interstellar clouds containing optically thin H2.
{\it Astrophys.J.}\,{\bf 197}, 575-580 \,(1975).\\
$^{18}$ Putman M.E., Gibson B.K. First results from the Parkes Multibeam High-velocity Cloud Survey.
{\it Proc.Astron.Soc.Australia}\,{\bf Vol.16, 1}, 70-76\,(1999).\\
$^{19}$ Grewing M., Schulz-L\"upertz E. Ionisationsstrukturen im galaktischen Halogas.\\
{\it Mitteil.Astron.Gesellsch.}\,{\bf 52},79-83\,(1981).\\
$^{20}$ de\,Boer K.S., Jura M.A., Shull J.M. Diffuse and dark clouds in the interstellar
medium. {\it Scientific Accomplishments of the IUE}. Ed. Y.Kondo. D. Reidel 
Publishing company. 485-515\,(1987).\\
$^{21}$ Morton D.C., Dinerstein H.L. Interstellar molecular hydrogen toward Zeta Puppis.\\
{\it Astrophys.J.}\,{\bf 204}, 1-11\,(1975).\\
$^{22}$ Akeson R.L., Blitz L. A search for hydrogen and molecular gas absorption in high-velocity
clouds. {\it Astrophys.J.}\,{\bf 523}, 163-170\,(1999).\\
$^{23}$ Pei Y.C., Fall S.M. Cosmic chemical evolution. {\it Astrophys.J.}\,{\bf 454}, 69-76\,(1995).\\

\newpage

\noindent
{\bf Acknowledgements}\\
\\
We thank M. Putman and C. Br\"uns for providing us with part of
the HIPASS data.\\
\\
\noindent
Correspondence and requests for material should be adressed to P.R. 
(email: prichter@astro.uni-bonn.de).

%*****
%\newpage

\begin{table}[h!]
\begin{tabular}{lrcl}
\multicolumn{4}{c}{\rule [-2mm]{0mm}{6mm}{Table 1: H$_2$ equivalent widths for the HVC gas towards HD\,269546}}
\vspace{0.02cm}\\
\\
\rule [-2mm]{0mm}{6mm}{Line} & $\lambda$ [\AA] & log $f \lambda ^{21}$ & $W$ [m\AA] \\
\vspace{0.02cm}\\
L\,R(0),4-0 &  $1049.366$ & $1.39$ & $\le 30$\\
L\,R(1),8-0 &  $1002.457$ & $1.26$ & $34 \pm 19$  \\
W\,Q(1),1-0 &  $986.798$  & $1.56$ & $75 \pm 22$  \\
W\,Q(2),0-0 &  $1010.941$ & $1.38$ & $65 \pm 22$  \\
L\,R(2),4-0 &  $1051.497$ & $1.19$ & $54 \pm 23$ \\
L\,P(2),7-0 &  $1016.472$ & $1.01$ & $\le 57$\\
L\,P(2),11-0&  $975.343$  & $0.81$ & $\le 42$\\
L\,P(3),3-0 &  $1070.142$ & $0.89$ & $33 \pm 17$  \\
L\,R(3),4-0 &  $1053.976$ & $1.18$ & $47 \pm 23$ \\
L\,R(3),8-0 &  $1006.418$ & $1.19$ & $63 \pm 30$  \\
L\,R(4),4-0 &  $1057.379$ & $1.17$ & $46 \pm 23$  \\
L\,P(5),4-0 &  $1065.594$ & $1.00$ & $\le 31$\\
L\,P(6),3-0 &  $1085.382$ & $0.94$ & $\le 54$ \\
L\,R(6),7-0 &  $1030.064$ & $1.17$ & $65 \pm 25$ \\
&\\
&\\
&\\

\end{tabular}
\end{table}

%*****
\newpage

\noindent
Fig.\,1: The spectral region between 1006 and 1012 \AA\, in the ORFEUS
spectrum of HD\,269546. The solid markings indicate H$_2$ absorption
lines with the Galactic component at 0 km\,s$^{-1}$ and
the arrows the HVC component at a Doppler velocity of $\sim$ +120 km\,s$^{-1}$.
Other identified absorption lines, which are of no
importance here, are marked with dashed lines.
\bigskip
\bigskip

\noindent
Fig.\,2: Comparison between the H$_2$ absorption profile (upper panel) and
the H\,{\sc i} emission profile (lower panel) for the HVC gas in direction
of HD\,269546.
The H$_2$ Q(2),0-0 line at 1010.941 \AA\, in the Werner band
(here plotted in the Doppler velocity scale, LSR)
reveals H$_2$ absorption from HVC gas near +120 km\,s$^{-1}$ (upper panel).
At this velocity, emission from neutral hydrogen$^{13}$ (lower panel)
is present, too.
\bigskip
\bigskip

\noindent
Fig.\,3: Map of the peak intensities (J2000) for neutral hydrogen emission from the
HIPASS survey
(preliminary, uncalibrated)
for velocities $v_{\rm LSR}$ = 105 - 145 km\,s$^{-1}$ in the field centred on HD
\,269546.
The map displays only the high intensities (dark structures).
The line of sight towards HD\,269546 passes between
two denser H\,{\sc i} clumps in the high-velocity cloud.
The line of sight to the LMC star HD\,36402,
on which absorption and emission is known at the
same velocity$^{10}$,
is also indicated.
The map was made available by the HIPASS team$^{18}$.

%*****
\newpage
\begin{figure}[t]
\resizebox{0.8\hsize}{!}{\includegraphics{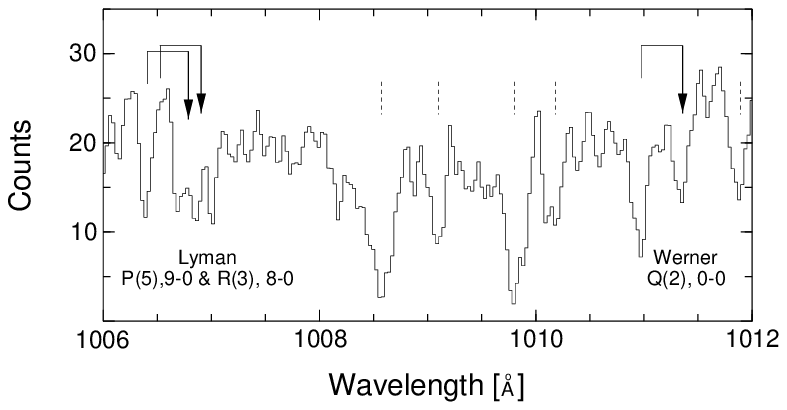}}

\vspace{7cm}
\end{figure}

\vspace{5cm}

%*****
\newpage
\begin{figure}[t]
\resizebox{0.5\hsize}{!}{\includegraphics{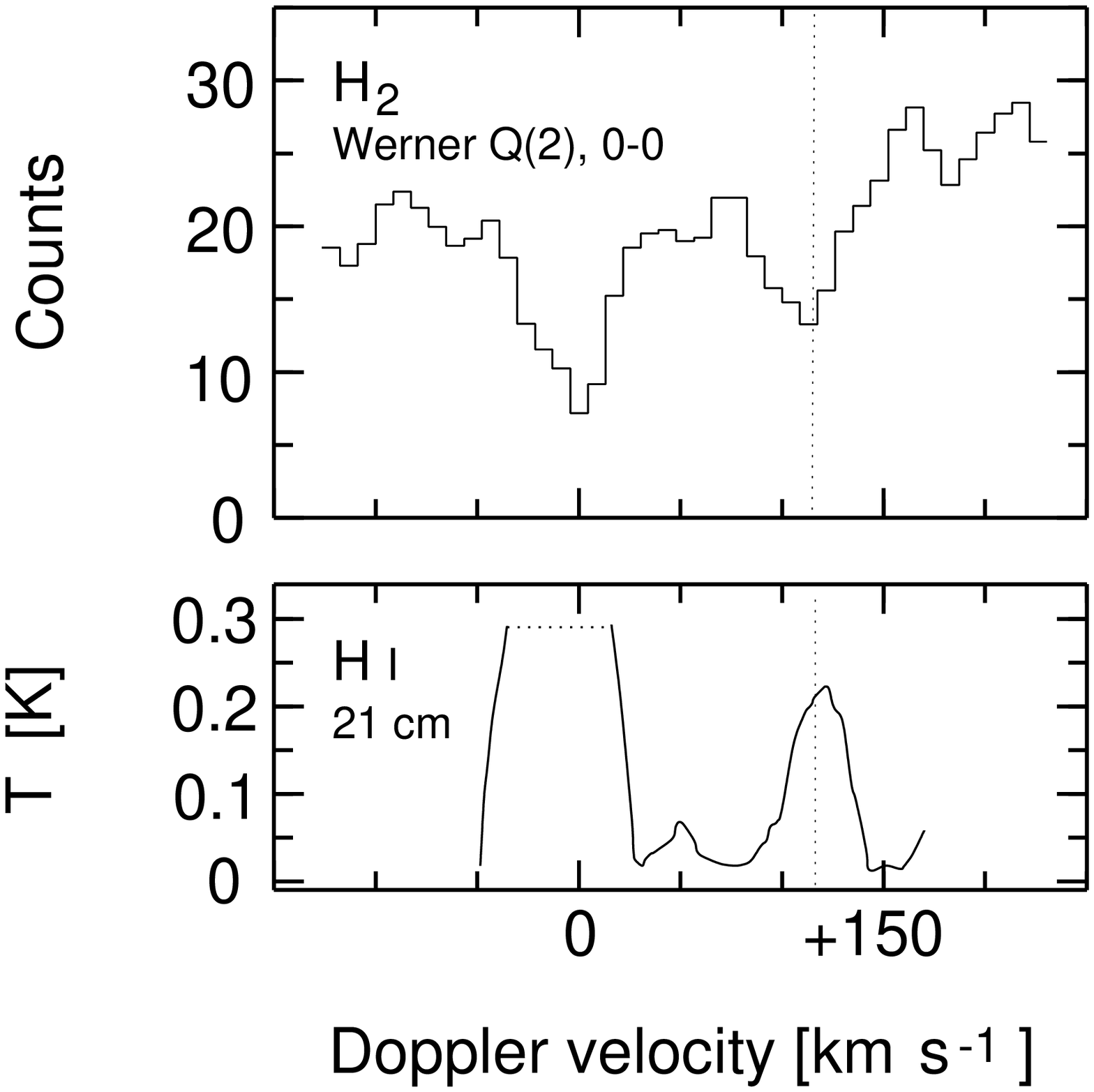}}

\vspace{7cm}
\end{figure}

%*****
\newpage
\begin{figure}
\resizebox{0.6\hsize}{!}{\includegraphics{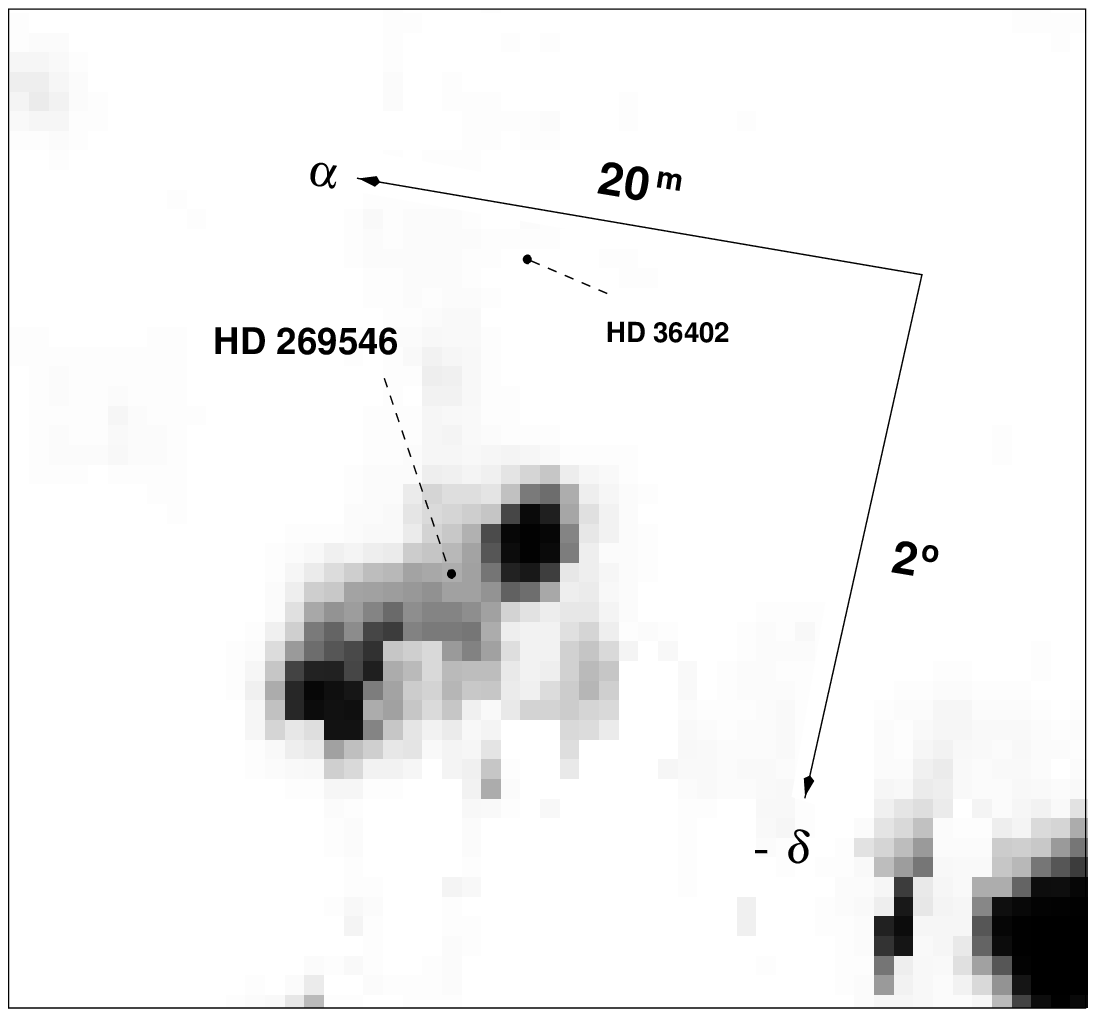}}
\end{figure}

\end{document}